\documentclass[journal]{IEEEtran}
\usepackage{graphicx}
\usepackage{epstopdf}
\usepackage{algorithm}
\usepackage{mathtools}
\usepackage{algorithmic}
\usepackage{amsfonts}
\usepackage{amssymb}
\usepackage{caption}
\usepackage{amsmath}
\input{Qcircuit}

\newcommand{\CASE}[1]{\STATE \textbf{case} #1\textbf{:} \begin{ALC@g}}
\newcommand{\ENDCASE}{\end{ALC@g}}

\newcommand{\DEFAULT}{\STATE \textbf{default:} \begin{ALC@g}}
\newcommand{\ENDDEFAULT}{\end{ALC@g}}
\newcommand{\DEFAULTLINE}[1]{\STATE \textbf{default:} }
\newcommand{\comment}[1]{}

\newtheorem{theorem}{Theorem}

\newtheorem{definition}[theorem]{Definition}

\begin{document}
\title{Optimizing Teleportation Cost in Distributed Quantum Circuits}
\author{Mariam~Zomorodi-Moghadam,
        Monireh~Houshmand,
        and~Mahboobeh~Houshmand
\thanks{M. Zomorodi-Moghadam is with the Department of Computer Engineering, Ferdowsi University of Mashhad, Mashhad, Iran, e-mail: (m$\_$zomorodi@um.ac.ir).}
\thanks{M. Houshmand is with the Department of Electrical Engineering, Imam Reza International University, Mashhad, Iran, email: (m.hooshmand@imamreza.ac.ir)}
\thanks{M. Houshmand is with the Department of Computer Engineering, Mashhad Branch, Islamic Azad University, Mashhad, Iran, email: (houshmand@mshdiau.ac.ir).}
\thanks{}}


\maketitle
\begin{abstract}
The presented work provides a procedure for optimizing the communication cost of a distributed quantum circuit (DQC) in terms of the number of qubit teleportations.
Because of technology limitations which do not allow large quantum computers to work as a single processing element, distributed quantum computation is an appropriate solution to overcome this difficulty.
Previous studies have applied ad-hoc solutions to distribute a quantum system for special cases and applications. In this study, a general approach is proposed to optimize the number of teleportations for a DQC consisting of two spatially separated and long-distance quantum subsystems.
  To this end, different configurations of locations for executing gates whose qubits are in distinct subsystems are considered and for each of these configurations, the proposed algorithm is run to find the minimum number of required teleportations. Finally, the configuration which leads to the minimum number of teleportations is reported. The proposed method can be used as an automated procedure to find the configuration with the optimal communication cost for the DQC.
\end{abstract}
\begin{IEEEkeywords}
Quantum Computation, Distributed Quantum Circuit, Teleportation Cost, Optimization
\end{IEEEkeywords}

\IEEEpeerreviewmaketitle

\section{Introduction}
Quantum computation~\cite{Nielsen10} is an interesting field of study with great potential to outperform classical computers in certain problems such as integer factorization~\cite{Grover}, discrete logarithm finding, and database search~\cite{shor-1997-26}.

Despite the theoretical advantages of quantum computations, realization of a large-scale quantum system is a real challenge~\cite{steffen2013}. It is mainly because that quantum computation technologies have limitations in the number of qubits they can process~\cite{fujii2012} and building a large-capacity quantum computer is hard~\cite{ying2009} which makes distributed implementations of quantum technologies necessary~\cite{van2010distributed}.

One of the limitations of quantum implementations is due to the interactions of qubits with the environment that leads to the decoherence and when the number of qubits increases, the quantum information becomes more fragile and more susceptible to errors~\cite{Krojanski2004}. To overcome the mentioned problems, using distributed nodes or subsystems in a quantum system is a reasonable solution in which fewer qubits are used in each node or subsystem. Therefore, to have a large quantum computer, one appropriate solution is to build a network of limited-capacity quantum computers which are interconnected through a quantum or classical channel and altogether can implement the behavior of the whole quantum system~\cite{Nickerson2013}. This structure is known as~\emph{distributed quantum computer}.

In order to perform a distributed quantum computation, there should be a reliable mechanism for communication between separate nodes of a distributed quantum system. Long-distance quantum communication is a technological challenge for physical realizations of quantum communication~\cite{Bras2008}. In this regard, teleportation~\cite{Bennett1993} is a primitive protocol for such this communication that uses entangled qubits to distribute quantum information~\cite{Whit2000}. This protocol has been experimentally implemented in many quantum technologies such as quantum photonics~\cite{bouwmeester1997}, NMR~\cite{Nielsen1998}, and trapped ions~\cite{Riebe2004}.

In teleportation, qubit states are teleported from one node to another, without physically moving them and then computations are performed locally on qubits, which is also known as \textit{teledata}. There is an alternative approach, called \textit{telegate} that executes gates remotely using the teleported gate approach without the need for qubits to be nearby~\cite{meter12008}. For some applications, like some adder algorithms it has been shown that teledata outperforms telegate and also its performance is independent of the network topology~\cite{Meter32006}.
In this study, teledata is considered for communication between distributed nodes.

Teleportation is a costly operation in a distributed quantum system. Moreover, according to no-cloning theorem~\cite{wootteres1982}, when a qubit is teleported to a destination subsystem, it can no longer be used in its own subsystem. Therefore, minimizing the number of teleportationn in a distributed quantum system is an important concern. In this study, this problem is considered and an algorithm is presented to optimize the number of teleportations for a distributed quantum system consisting of two spatially separated and long-distance quantum subsystems.

The paper is organized as follows: Section 2 explains some basic concepts and definitions of quantum computation and distributed quantum computing. Related work is presented in Section 3. Section 4 then establishes some definitions and notation needed for the proposed algorithm. Our proposed algorithm for optimal distributed quantum circuit is described in Section 5. Finally in Section 6, we conclude the paper.

\section{background}
Quantum bits or \emph{qubits} are quantum analogues of classical bits. A qubit is a two-level quantum system whose state is represented by a unit vector in a two-dimensional Hilbert space, $\mathcal{H}_2$, for which an orthonormal basis, denoted by $\{$$\left\vert 0\right\rangle$, $\left\vert 1\right\rangle$$\}$, has been fixed. Unlike classical bits, qubits can be in a superposition of $\left\vert 0\right\rangle$ and $\left\vert 1\right\rangle$ which is represented as $\alpha\left\vert 0\right\rangle+\beta\left\vert 1\right\rangle$, where $\alpha$ and $\beta$ are complex numbers such that
$|\alpha|^2 + |\beta|^2 = 1$. If such a superposition is measured with respect to the $\{$$\left\vert 0\right\rangle$, $\left\vert 1\right\rangle$$\}$ basis, then the classical outcome of 0 is observed with the probability of $|\alpha|^2$ and the classical result of 1 is observed with the probability of $|\beta|^2$. If 0 is obtained, the state of the system after measurement will collapse to $\left\vert 0\right\rangle$ and if 1 is obtained, it will be $\left\vert 1\right\rangle$.

There are a number of models for the evolution of quantum computation. The main model to explore quantum
computation is the circuit model, based on unitary evolution of qubits by networks of gates~\cite{Nielsen10}.
Every $n$-qubit quantum gate is a linear transformation represented by a unitary matrix defined on an $n$-qubit Hilbert space. A matrix $U$ is \emph{unitary} if $UU^{\dagger} = I$, where $U^{\dagger}$ is the conjugate transpose of the matrix $U$.

Some useful single-qubit gates are the elements of the Pauli set:
\[
\sigma_0=
I=%
\begin{bmatrix}
1 & 0\\
0 & 1
\end{bmatrix},
\sigma_1=
X=%
\begin{bmatrix}
0 & 1\\
1 & 0
\end{bmatrix}, \]
\[
\sigma_2=
Y=%
\begin{bmatrix}
0 & -i\\
i & 0
\end{bmatrix},
\sigma_3=
Z=%
\begin{bmatrix}
1 & 0\\
0 & -1
\end{bmatrix}.
\]

Another class of useful unitary gates on a single qubit are rotation operators around $x$, $y$ and $z$ axis with the angle $\alpha$ in the Bloch sphere, as shown below:

\[
R_x (\alpha )=%
\begin{bmatrix}
{{\rm cos}\frac{\alpha }{{\rm 2}}} & { - i{\rm sin}\frac{\alpha }{{\rm 2}}}\\
{- i{\rm sin}\frac{\alpha }{{\rm 2}}} & {{\rm cos}\frac{\alpha }{{\rm 2}}}
\end{bmatrix},
R_y (\alpha )=%
\begin{bmatrix}
{{\rm cos}\frac{\alpha }{{\rm 2}}} & { - {\rm sin}\frac{\alpha }{{\rm 2}}}\\
{{\rm sin}\frac{\alpha }{{\rm 2}}} & {{\rm cos}\frac{\alpha }{{\rm 2}}}
\end{bmatrix}\]
\[
R_z (\alpha )=%
\begin{bmatrix}
{e^{ - i\frac{\alpha }{2}} } & 0\\
0 & {e^{ i\frac{\alpha }{2}} }
\end{bmatrix}.\]

Hadamard, $H$, and \emph{T} are two other known single-qubit gates where:
\[
H=%
\frac{1}{\sqrt{2}}\begin{bmatrix}
1 & 1\\
1 & -1
\end{bmatrix},
T=
\begin{bmatrix}
1 & 0\\
0 & e^{i\frac{\pi }{4}}\,
\end{bmatrix}
.
\]

\begin{figure}
\centering
\[
\Qcircuit @C=1.0em @R=1.4em{
& \ctrl{1}&\qw\\
& \targ&\qw
}
\]
\caption{The circuit representation of CNOT gate.}
\label{fig:cnot}
\end{figure}
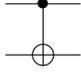

If $U$ is a gate that operates on a single qubit, then controlled-$U$ is a gate that operates on two qubits, i.e., control and target qubits, and \emph{U} is applied to the target qubit if the control qubit is $\left\vert 1\right\rangle$ and leaves it unchanged otherwise. For example, controlled-NOT (CNOT) gate performs the $X$ operator on the target qubit if the control qubit is $\left\vert 1\right\rangle$. Otherwise, the target qubit remains unchanged.
Figure~\ref{fig:cnot} shows the circuit representation of the CNOT gate.

 To realize arbitrary quantum gates, they are decomposed to a set of physically implementable gates by quantum technologies, which is called quantum logic synthesis~\cite{Mottonen-2004,Shende06,houshmand2014,houshmand2015}. This set of gates typically consists of CNOT and single-qubit gates, called ``basic gate" library~\cite{Barenco95} or CNOT and single-qubit rotation gates, called ``elementary gate" library~\cite{bullock2004}.

A quantum circuit consists of quantum gates interconnected by quantum wires carrying qubits with time flowing from left or right. The unitary matrix of the quantum circuit is evaluated by either dot product or tensor product of the unitary matrices of those quantum gates. The net effect of the gates which are applied to a same subset of qubits in series is computed by the dot product which is the same as the known matrix multiplication. The adjacent gates which act on independent subsets of qubits can be applied in parallel and their overall net effect is computed by their tensor product and defined as follows.

Let $A$ be an $m\times n$ matrix and let $B$ be a $p\times q $ matrix. Then $A \otimes B$ is an $(mp)\times(nq)$ matrix called the tensor product (Kronecker product) of $A$ and $B$ and is defined as below:
\[
A \otimes B = \left[ \begin{array}{l}
 a_{11} B,a_{12} B,...,a_{1n} B \\
 a_{21} B,a_{22} B,...,a_{2n} B \\
 \,\,\,\,\,\,\,\,\,\,\,\,\,\,\,\,\,\,... \\
 a_{m1} B,a_{m2} B,...,a_{mn} B \\
 \end{array} \right]
\]
where $a_{ij}$ shows the element in the $i^{th}$ column and the $j^{th}$ row of matrix $A$.

\emph{Entanglement} is a quantum mechanical phenomenon that plays a key role in many of the most interesting applications of quantum computation and quantum information.
A multi-qubit quantum state $\left\vert \psi \right\rangle$ is said to be entangled if it cannot be written as
the tensor product $\left\vert \psi \right\rangle=\left\vert \phi_1 \right\rangle \otimes \left\vert \phi_2\right \rangle $ of two pure states. For example, the EPR pair~\cite{Nielsen10} shown below is an entangled quantum state:
\[\left\vert \Phi  \right\rangle=(\left\vert 00 \right\rangle+\left\vert 11 \right\rangle)/\surd{2}\]

A distributed quantum circuit ($DQC$) is an extension to the quantum circuit model.
 To have a distributed quantum computing system, the quantum states have to be moved between computing nodes.
 Teleportation is a known technique for communication between distributed nodes.

\begin{figure}

\centering
\[
\Qcircuit @C=1em @R=1em{
\lstick{\ket{\psi}} & \ctrl{1} & \gate{H} &\meter &\ustick{M_1} \cw  & \cw & \cw & \cw & \control \cw & & \\
 & \targ & \qw & \meter & \ustick{M_2} \cw & \cw &  \control \cw & &\cwx & & \\
\lstick{\ket{\beta_{00}}} & & & & & & \cwx & &\cwx & & \\
 & \qw &\qw & \qw & \qw & \qw & \gate{X^{M_2}} \cwx & \qw & \gate{Z^{M_1}} \cwx & \qw & \rstick{\ket{\psi}} \gategroup{2}{1}{4}{1}{.5em}{\{}
}
\]
\caption{Quantum circuit for teleporting a qubit~\cite{Nielsen10}.}
\label{fig:tele}
\end{figure}
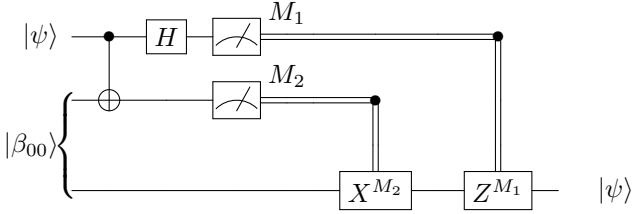

Teleportation allows us to send the state of a qubit from point $A$ to point $B$ by communicating only two classical bits. In this study, it is assumed that data transfer between distributed nodes in the system is performed by teleportation. Figure~\ref{fig:tele} shows the quantum circuit for a basic quantum teleportation as described in~\cite{Nielsen10}. In this figure, the two top lines represent the sender's system and the bottom line is the receiver's one. The meters represent measurement, and the double lines coming out of them carry classical bits.

As stated in \cite{Bennett1993} and mathematically proved in \cite{ying2009}, it can be concluded that teleporting a quantum state works even if that state is a mixed or an entangled state and so it can be used as communication protocol between distributed quantum systems.

\section{Related Work}

Distributed quantum computing has been studied for more than fifteen years~\cite{ying2009} and the first suggestions were made by Grover \cite{Grover1997}, Cleve and Buhrman~\cite{Cleve1997} and later by Cirac et.al., \cite{Cirac1999}.
In \cite{Grover1997}, Grover presented a distributed quantum system where there are some particles at remote locations and each one performs its computation and sends the required information to a base station when necessary.
He mentioned that the issue in distributed quantum computing is how to optimally divide the problem into the different quantum computing units. The answer to this question effects the performance of the proposed problem in this work that can be considered in the future.
Grover showed that using this distributed approach, the overall computation time is faster proportional to the number of such distributed particles. In the same paper, he also presented a quantum algorithm and adapted it to distributed computation where communication is costly.

Beals et.al., \cite{Beals2013} assumed a distributed quantum computer with nodes connected according to the hypercube graph and used a few long-range qubits to emulate a quantum circuit with a reasonable overhead. They showed that an arbitrary quantum circuit can be emulated by a distributed quantum circuit with nodes connected using a hypercube graph.

Yepez~\cite{yepez2001type} presented an architecture for distributed quantum computing with two types of communication which were called type I and type II. Type I quantum computer uses quantum communication between subsystems and type II quantum computer exploits classical communication between subsystems of the distributed computer. Using this classification, our distributed quantum circuit is of type I.\\
Related to teleportation cost, Streltsov et.al.,~\cite {Streltsov2012} posed the question of the cheapest way for distributing entanglement and provided the minimal quantum cost for sending an entangled composite state in a long distance. They showed that regarding the most general distribution protocol, the amount of entanglement sent in the total process of distributed communication cannot be larger than the total entanglement cost for sending the ancilla particle and sending back that particle.

Classical communication cost is the subject of the study in \cite{lo2000classical}.
Authors conjecture that in a two-stage teleportation, each step requires a single bit of classical communication and in general, for an arbitrary $N$-dimensional pure state, $2log_2N$ bits of classical communication is required for remote preparation which is different from the usual teleportation in which the precise state of the qubit to be prepared in the receiver is known to the sender.

Ying and Feng \cite{ying2009} provided some definitions of a distributed quantum computing system and defined an algebraic language to describe quantum circuits for distributed quantum computing. Van Meter et.al.,~\cite{meter12008} designed a fixed distributed quantum circuit for VBE carry-ripple adder. They evaluated the cost of the teleportation for the designed distributed system. They compared two approaches, namely teledata and telegate topologies of quantum circuits and showed that teleportation of data (the one we use in our work) is better than teleportation of gates. They also compared it to a monolithic machine and showed that as the node sizes are increased the multicomputer architecture has better performance in large problem sizes. But their distributed quantum circuit was a predesigned architecture and they didn't make any specific change in the system to reduce the number of teleportation circuits in the design.

Also in \cite{Yimsiriwattana2004}, authors used the teleportation communication model for implementing a distributed version of Shor's algorithm. But like the one in \cite{Meter2006}, their method for assigning logical qubits to distributed nodes is an ad-hoc method specific for implementing Shor's algorithm. In their implementation, each computing node holds $n/4$ qubits from each register and output of each computing node is teleported to the next computing node. So there is no optimization in the number of teleportations and also the authors did not consider the cost of returning the teleported qubits back into their own subsystems.

\section{Definitions and notations}
\label{sec:def}

In this section, we establish some assumptions and notations before proceeding with the main development of the study.
First of all, we assume that there is a quantum circuit, $QC$ as input with width $W$, size $S$, and depth $D$ with the following definitions as stated in \cite{pham20132d}:

The \emph{width} of a quantum circuit is defined as the total number of qubits in the circuit.

The \emph{size} of a quantum circuit is the total number of its gates from a set of universal quantum gates.

The \emph{depth} of a quantum circuit is the number of timesteps, $D$ required for executing the circuit. In each timestep one or more gates can be executed in parallel.

It is assumed that there is an appropriate ordering of gate executions so that the number of timesteps is minimal.
In each timestep, a set of gates that can be performed in parallel are executed.

\begin{definition}
A \emph{Distributed Quantum Circuit (DQC)} consists of $N$ limited capacity \emph{Quantum Circuits (QCs)} or partitions which are located far from each other and altogether emulate the functionality of a large quantum circuit.
Partitions of a $DQC$ communicate by sending their qubits to each other using a specific quantum communication channel through teleportation.
\end{definition}

For simplicity, in this paper it is assumed that there are just two partitions in $DQC$, each with the size of $W/2$.
Without losing the generality, $W$ is assumed to be an even number.
Qubits are numbered from top to bottom from one to $n$ in each partition, where the $i^{th}$ line of the circuit from top to bottom represents the $i^{th}$ qubit, \textit{$q_i$}.

The gates are numbered in order of their executions in the $QC$ and $g_i$ means the $i^{th}$ gate which is executed according to a scheduling algorithm. For gates which can be executed in parallel, the priority of their numbering is arbitrary. The set of all gates in the circuit is represented by $\mathcal{G}$.

There are three kinds of quantum gates in the $DQC$, namely, single-qubit, local and global CNOT gates and their definitions and representations are as follows:
\begin{itemize}
\item
A single-qubit gates is shown as $gate\_name_i (p, j)$. $p,$ indicates the index of partition to which the gate $gate\_name_i$ belongs and $j$ indicates the index of the qubit the gate acts on.
\item
A \emph{local} CNOT gate is the one whose control and target qubits belong to the same partition and is shown as $\text{CNOT}_i(p, j_c, j_t)$.
$p$ indicates the index of partition to which the gate $\text{CNOT}_i$ belongs. $j_c (j_t)$ indicates the index of control (target) qubit in its partition.
\item
A \emph{global} CNOT gate is the one whose control and target qubits belong to different partitions is called a \emph{global} CNOT gate shown as $\text{CNOT}_i(p_c, j_c, p_t, j_t)$.
$j_c (j_t)$ indicates the index of control (target) qubit in its home partition.
The partition to which each qubit $q$ of a global CNOT gate belongs is called the \emph{home} partition of $q$. $p_c (p_t)$ indicates the index of partition to which the control (target) of $\text{CNOT}_i$ belongs.
\end{itemize}

It is supposed that local gates including single-qubit and local CNOT gates are performed in their local partitions. The total number of gates in a $QC$ and the number of global gates are denoted by $m_t$ and $m_g$ respectively. Also a subset of $\mathcal{G}$ for representing the set of global gates is denoted by  $\mathcal{G}_d$.

\emph{$Config$-$Arr$}, is an array with the size of $m_g$, whose elements show the partitions each global gate is supposed to be executed.
The value of `0' (`1') for the elements of this array means the corresponding gate is assumed to be executed in $P_0$ ($P_1$). The partition in which the gate $g$ is supposed to be executed based on the {$Config$-$Arr$} array is denoted by $g.l$.

\begin{definition}
In order to perform a global CNOT gate, one of its two qubits should be teleported from its home partition to another.
This qubit is called a \emph{migrated} qubit, as long as it is not teleported back to its home partition.
\end{definition}

To keep load balancing of quantum subsystems, it is assumed that at each time, the number of migrated qubits is at most one.

For example, in the circuit of Figure~\ref{fig:QC1}, the set $\mathcal{G}$ is represented as:\\
$\{\text{CNOT}_1(0,0,1)\,\text{CNOT}_2(1,0,0,0)\,\text{CNOT}_3(0,0,1,1)\,H_4(1,1)$
$\text{CNOT}_5(1,1,0,1)\,H_6(0,1)\,\text{CNOT}_7(0,1,1,1)$
$\text{CNOT}_8(0,0,1)\,\text{CNOT}_9(1,1,0,0)\}$

\begin{figure}[!htb]
\begin{center}
\includegraphics[width=0.5\textwidth]{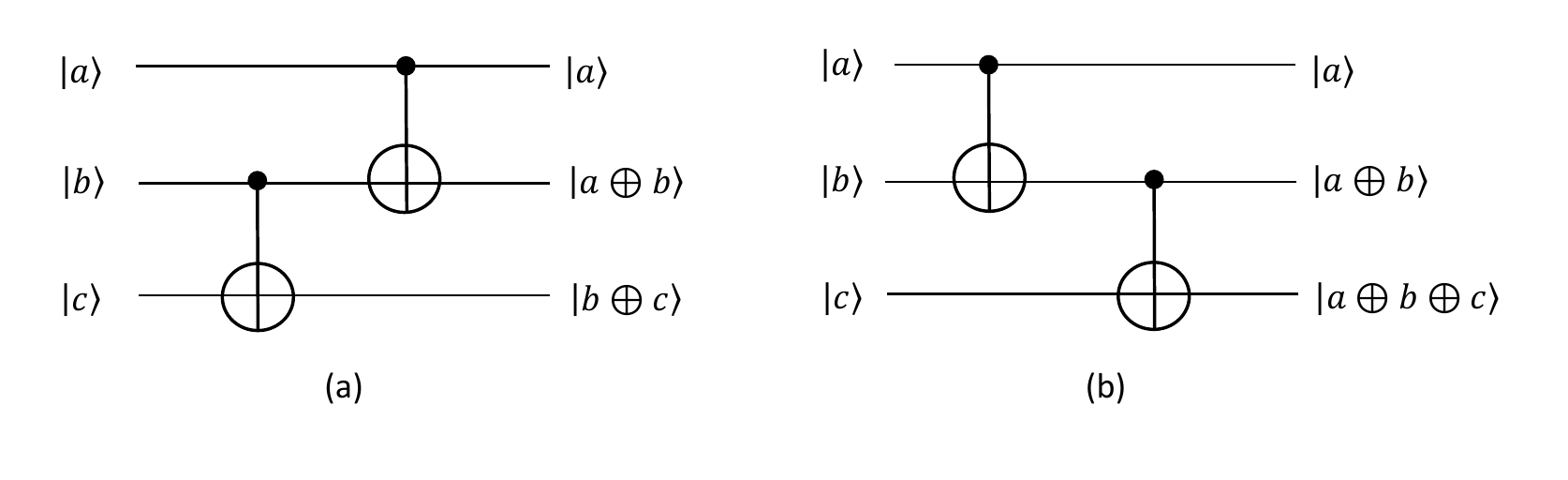}
\end{center}
\caption{(a) and (b) show when the index of the control qubit of one CNOT is the same as the index of the target of the other, two CNOTs do not commute~\cite{houshmand2012minimal}.}
\label{fig:commute}%
\end{figure}

As non-commutativity property of gates is an important concept in our algorithm, it is considered in the following.
First we examine the non-commutativity of two CNOT gates. Two CNOT gates do not commute when the index of
control qubit of CNOT gate is the same as the index of target qubit of other gate~\cite{houshmand2012minimal} as shown in Figure~\ref{fig:commute}. As shown in this figure, the third output in circuit in the left is $\left| {b \oplus c} \right\rangle$, while it is $\left| {a \oplus b \oplus c} \right\rangle$ in the circuit in the right.

Now the commutativity relations of two adjacent CNOT and a single-qubit gate are considered.
 \begin{enumerate}
  \item  If the single-qubit gate, $U$ is applied to the same qubit of control of the CNOT gate, they commute if:
\begin{equation}
\label{eq1}
(U \otimes I)\text{CNOT}=e^{i\theta}\text{CNOT} (U \otimes I)
\end{equation}
where $\theta \in [0,2\pi).$
Examining~(\ref{eq1}), $\theta$ is obtained 0, and $U$ is a diagonal unitary matrix as follows:
\[U=
\begin{bmatrix}
u_{0} &0\\
0 & u_{1}
\end{bmatrix}\].

\item If the single-qubit gate, $U$ is applied to the target qubit of CNOT gate they commute if that the $X$ gate and $U$ gate commute.
$X$ gate and a $U$ gate
commute if:

\begin{equation}
\label{eq2}
XU=e^{i\theta}UX,
\end{equation}

Examining~(\ref{eq2}), we found that $\theta$ is either 0 or $\pi$. In the case of $\theta=0$, the structure of $U$ is a unitary matrix as follows:
\[U=\begin{bmatrix}
u_{0} & u_{1}\\
u_{1} & u_{0}
\end{bmatrix},\]
and where $\theta=\pi,$ the structure of $U$ is a unitary matrix as follows:
\[U=\begin{bmatrix}
u_{0} & u_{1}\\
-u_{1} & -u_{0}
\end{bmatrix}.\]

\end{enumerate}

\section{proposed approach}
\label{sec:2}
In this section, first the proposed algorithm for optimizing the teleportation cost of a given $DQC$ is explained and then, it is performed on a simple running example.
\subsection{The proposed algorithm}
\label{sec:main}
In this section, the proposed algorithm to find the optimal number of teleportations and their sequence for a given $DQC$ is presented. The algorithm is motivated by the fact that teleportation is a costly operation and reducing the number of teleportations in a given $DQC$ is of great importance.

We intend to start with a quantum circuit $QC$, composed of basic gate library, i.e., CNOT and single-qubit gates operating on $2n$ qubits. It is assumed that $QC$ is already split into two partitions, $P_0$ and $P_1$ each with $n$ qubits.

Suppose there are $m_g$ global CNOT gates in the given $DQC$. In general, each of these gates can be executed in either $P_0$ or $P_1$  and hence there are $2^{m_g}$ different configurations for executing $m_g$ global gates. In the proposed approach, all of these configurations are considered and for each of which, the minimum number of teleportation and their corresponding sequence are analytically determined. Finally, the configuration which leads to the minimum number of teleportation is reported. This procedure is performed by Algorithms~\ref{alg:main} and~\ref{alg:mint} as explained in the following.

Algorithm \ref{alg:main} is the main algorithm which receives the reported teleportation cost for each configuration from Algorithm \ref{alg:mint} and returns the configuration with the minimum cost as output.

Algorithm \ref{alg:mint} takes a $DQC$ with the ordered list of gates, $\mathcal{G}$, and $Config$-$Arr$ as inputs and returns the minimum number of teleportations and their sequence. This algorithm is first called with $n_t = 0$.  The subset of global gates, $\mathcal{G}_d$ is also used in the algorithm for reporting the sequence of teleportations. In different steps of this algorithm, the general and the global gates which are executed are removed from $\mathcal{G}$ and $\mathcal{G}_d$, respectively.

 This algorithm starts from the first gate in list $\mathcal{G}$ of $DQC$. If it is a local gate, then no teleportation is needed for it and this gate is removed from $\mathcal{G}$. Otherwise, the qubit is teleported to the destination and the teleportation cost ($n_t$) is increased by one. Furthermore, to show the sequence of teleportations, $\mathcal{G}_d[0](C \slash T)$ is reported. Based on $Config$-$Arr$, $C \slash T$ determines whether the control$\slash$ target qubit of $\mathcal{G}_d[0]$ should be telepoted to the other partition.

 When a qubit of a global gate, $temp$ in algorithm, is teleported to the other partition, the whole circuit is tracked and as much as possible number of gates that can be executed without the need of teleporting back the qubit are executed. This means that the migrated qubit has been used optimally by other gates before it is teleported back to its own partition.

 Three characteristics of gates that can not be executed are explained in the $Non$-$Execute$ function. This function returns TRUE when any of these characteristics occurs. When there is a gate, $\mathcal{G}[i]$, for which $Non$-$Execute(temp,\mathcal{G}[i])$ function returns FALSE, there are possibly some gates before $\mathcal{G}[i]$ which have not been executed. In this case, $\mathcal{G}[i]$ can be executed just in case that it can commute with all of these gates. The commutativity of gates is checked in the $Non$-$Commute$ function. The mentioned functions are introduced in the following:

\begin{itemize}
\item
$Non$-$Execute(g, g')$, takes two gates, $g$ and $g'$ as inputs. It returns TRUE, if the migrated qubit of $g$ should be returned to its home partition due to the three different cases mentioned below and then $g'$ can be performed. It returns FALSE otherwise.
\begin{enumerate}
\item $g'$ is a local gate whose one of qubits is the same as the migrated qubit of $g$.\\
\item $g'$ is a global gate with different label of $g$, i.e., $g'.l != g.l$.\\
\item $g'$ is a global gate and $g'.l = g.l$, but it requires another teleportation in order to execute $g'$.
\end{enumerate}

\item
$Non$-$Commute(g,g')$, takes two gates, $g$ and $g'$ as inputs. It returns TRUE, if the gates $g$ and $g'$ do not commute and returns FALSE otherwise. According to discussions in Section~\ref{sec:def}, the non-commutativity of two gates happens in three cases:\\
\begin{enumerate}
\item The two gates are CNOTs where the index of control qubit of one of them is the same as the index of target qubit of the other.\\
\item One of the gates is CNOT and the other is a non-diagonal single-qubit gate that acts on the control qubit of CNOT.\\
\item One of the gates is CNOT and the other is a single-qubit gate which acts on the target qubit of CNOT and does not have any of the structures as explained in (\ref{eq1}) or (\ref{eq2}).
\end{enumerate}

\end{itemize}

At the end of each run of the algorithm, $n_t$ is increased by one once more because another teleportation is required to return the teleported qubit back. Then the algorithm recursively is called with the new existing gates until there is no gate in the set $\mathcal{G}$ and the minimum number number of teleportations ($n_t$) is obtained.

\begin{algorithm}
\caption{Main algorithm}
\label{alg:main}
\begin{algorithmic}[1]
\FOR{$i\leftarrow 1$ \TO $2^m$ configurations of $Config$-$Arr$}
\STATE $n_t$=0
\STATE temp[i]=$Min$-$Teleportation(\mathcal{G}$,$Config$-$Arr$,$n_t$)
\ENDFOR
\STATE \textbf{return} min (temp)
\end{algorithmic}
\end{algorithm}

\begin{algorithm}
\caption{Algorithm for determining the minimum number of teleportations required in $\mathcal{G}$}
\label{alg:mint}
\begin{algorithmic}[1]
\STATE $Min$-$Teleportation(\mathcal{G}$,$Config$-$Arr$,$n_t$)
\STATE sw = 0
\IF{$Empty(\mathcal{G})$}
\STATE \textbf{return $n_t$}
\ENDIF
\IF{$local (\mathcal{G}[0])$}
\STATE \textbf{remove}($\mathcal{G}[0]$)
\STATE $Min$-$Teleportation(\mathcal{G}$,$Config$-$Arr$,$n_t$)
\ENDIF
\STATE $temp=\mathcal{G}[0]$
\STATE $n_t = n_t + 1$
\STATE \textbf{remove}($\mathcal{G}[0]$)
\STATE \textbf{report}($\mathcal{G}_d[0](C \slash T$))
\STATE \textbf{remove}($\mathcal{G}_d[0]$)
\FOR{$i\leftarrow0$ \TO $\mathcal{G}.length$}
 \IF{$Non$-$Execute(temp,\mathcal{G}[i])$=\textbf{FALSE}}
             \FOR{$k\leftarrow{i}$ \TO $0$}
        \IF{$Non$-$Commute(\mathcal{G}[i],\mathcal{G}[k])$=\textbf{TRUE}}
            \STATE sw = 1
			\STATE \textbf{break}
 \ENDIF
      \ENDFOR
\IF {sw = 0}
\STATE \textbf{remove}($\mathcal{G}[i]$)
 \IF{$global(\mathcal{G}[i]$)}
\STATE \textbf{remove}($\mathcal{G}[i]$ \textbf{from} $\mathcal{G}_d)$
\ENDIF
\ENDIF
 \ENDIF
       \ENDFOR
       \STATE $n_t = n_t + 1$
      \STATE $Min$-$Teleportation(\mathcal{G}$,$Config$-$Arr$,$n_t$)
\end{algorithmic}
\end{algorithm}

\subsection{Running Example}
\begin{figure}[!htb]
\begin{center}
\includegraphics[width=0.5\textwidth]{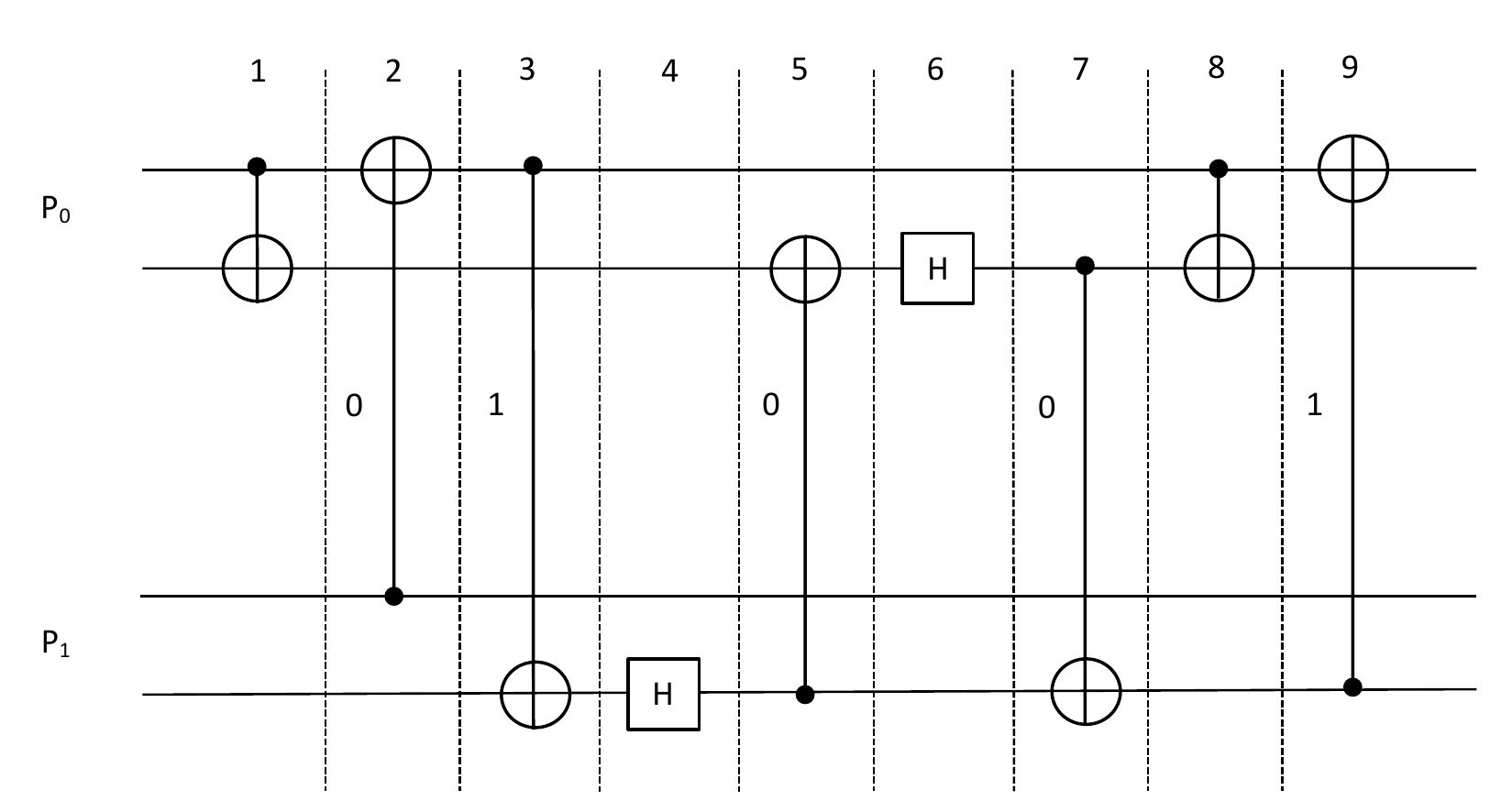}
\end{center}
\caption{A sample $DQC$}
\label{fig:QC1}%
\end{figure}

In this section, the proposed algorithm in Section~\ref{sec:main} is explained by a running example. This example is illustrated in Figure~\ref{fig:QC1} and the proposed algorithm is applied to it.

\begin{figure}[!htb]
\begin{center}
\includegraphics[width=0.5\textwidth]{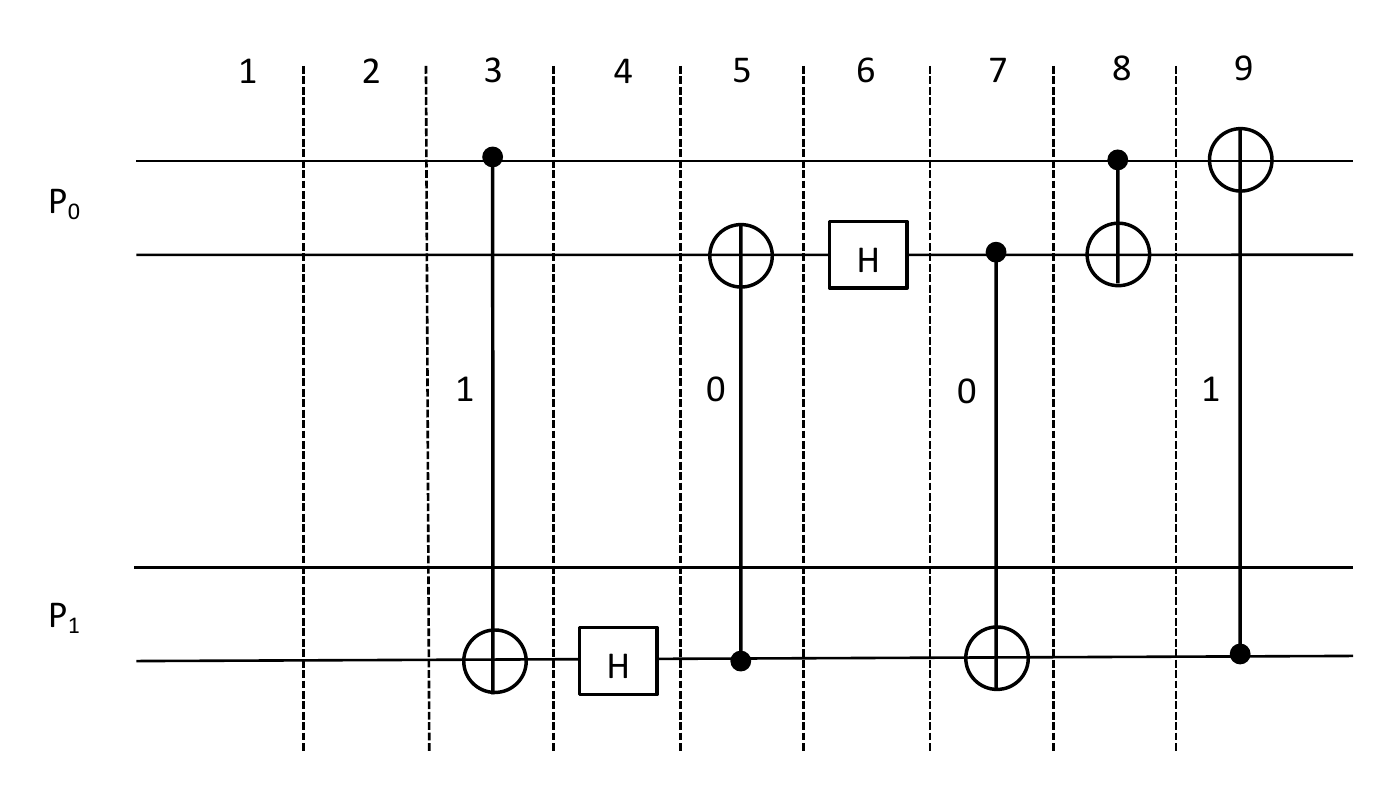}
\end{center}
\caption{New $DQC$ after the first run of Algorithm~\ref{alg:mint}}
\label{fig:QC2}%
\end{figure}

\begin{figure}[!htb]
\begin{center}
\includegraphics[width=0.5\textwidth]{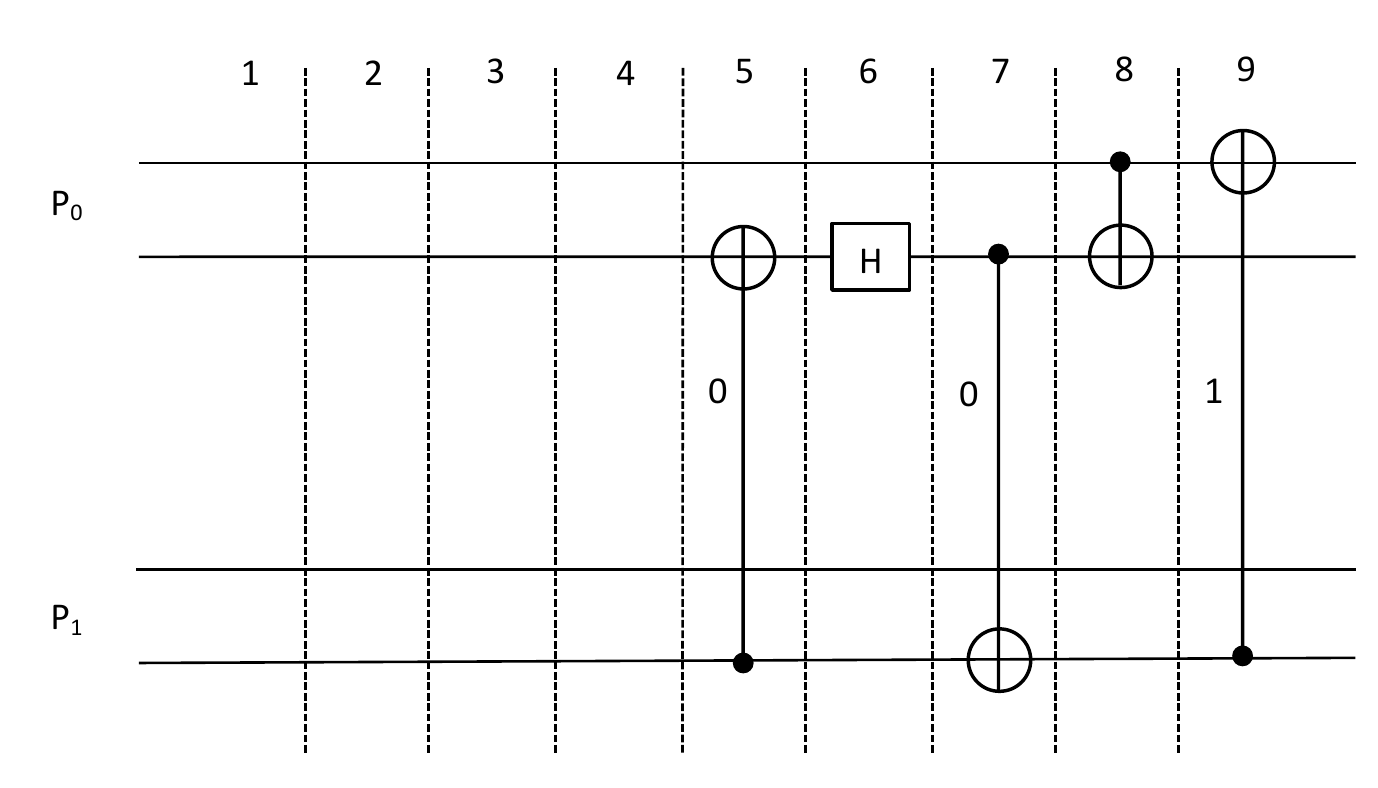}
\end{center}
\caption{New $DQC$ after the second run of Algorithm~\ref{alg:mint}}
\label{fig:QC3}
\end{figure}

\begin{figure}[!htb]
\begin{center}
\includegraphics[width=0.5\textwidth]{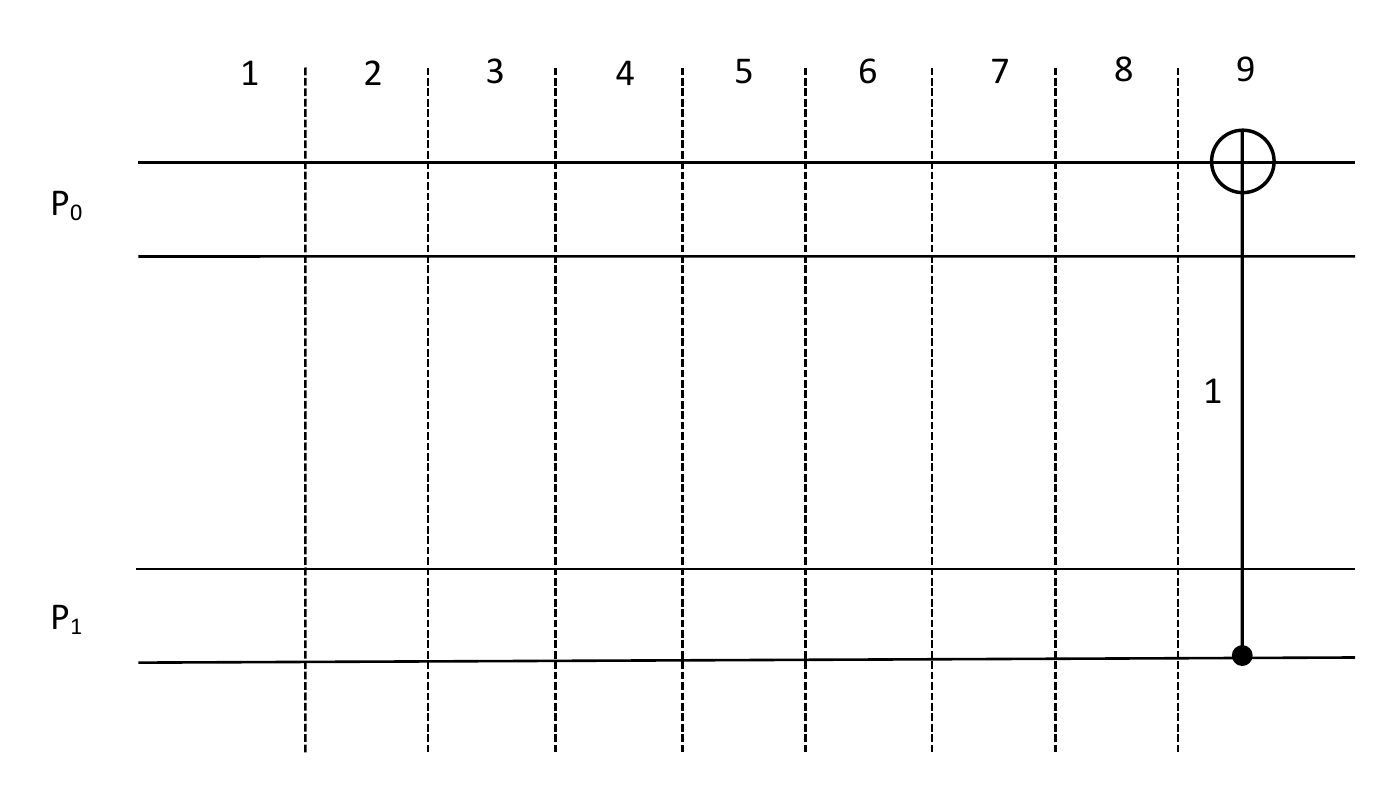}
\end{center}
\caption{New DQC after the third run of Algorithm~\ref{alg:mint}}
\label{fig:QC4}%
\end{figure}

The algorithm starts by taking an already two-partite $DQC$ as an input. This circuit consists of nine gates including two single-qubit gates, two local CNOT gates and five global CNOT gates.
Since there are five global gates, the $Config$-$Arr$ has five elements and $2^5=32$ different configurations. Algorithm~\ref{alg:main} for all these configurations calls Algorithm~\ref{alg:mint} which finds the minimum number of teleportation and their sequence for that configuration. Algorithm 1 finally reports the configuration with the smallest number of teleportations. Therefore, for the given $DQC$ this approach finds the minimum number of teleportations.

Let us consider $Config$-$Arr$=$\{0,1,0,0,1\}$ as a sample configuration. Algorithm~\ref{alg:mint} starts with the first gate in $\mathcal{G}$, i.e., $\text{CNOT}_1(0,0,1)$ which is a local gate. Therefore, it is executed in its own partition and is then removed. The next gate in this array is $\text{CNOT}_2(1,0,0,0)$ which is a global gate and based on the first element of $Config$-$Arr$, this gate is executed in $P_0$.
For executing this gate, qubit $\#0$ in $P_1$ is teleported to $P_0$, the number of teleportations, $n_t$ is increased by one and then this gate is removed.
In the outer loop of Algorithm 2 (Line 15), for all other gates, they are removed from the list, if they can be executed before teleporting back this migrated qubit.

The steps for the first run of Algorithm~\ref{alg:mint} are as follows.
\begin{itemize}
\item $Non$-$Execute (\text{CNOT}_2 (1,0,0,0), \text{CNOT}_3 (0,0,1,1))$= TRUE according to the second condition of $Non$-$Execute$ function as they are intended for execution in different partitions.

\item $Non$-$Execute(\text{CNOT}_2(1,0,0,0), H_4(1,1))$= FALSE. Therefore, in the inner loop of Algorithm~\ref{alg:mint} (Line 17), this gate is checked with the set of all previous existing gates whether they can commute. As $Non$-$Commute(\text{CNOT}_3(0,0,1,1),H_4(1,1))$= TRUE, $H_4(1,1)$ can not be executed and is not removed.

\item $Non$-$Execute(\text{CNOT}_2(1,0,0,0), \text{CNOT}_5(1,1,0,1))$= TRUE according to the third condition of $Non$-$Execute$ function as a different qubit should be teleported for the execution of this gate.

\item $Non$-$Execute(\text{CNOT}_2(1,0,0,0), H_6(0,1))$= FALSE. Therefore, the algorithm enters the inner loop where this gate is checked with the set of all previous existing gates for possible commutativity. In this case, $Non$-$Commute(\text{CNOT}_5(1,1,0,1),H_6(0,1))$= TRUE and hence, $H_6(0,1)$ cannot be executed and is not removed.

\item $Non$-$Execute(\text{CNOT}_2(1,0,0,0), \text{CNOT}_7(0,1,1,1))$= TRUE since the case three of $Non$-$Execute$ function is satisfied.

\item $Non$-$Execute(\text{CNOT}_2(1,0,0,0), \text{CNOT}_8(0,0,1))$= FALSE but as $Non$-$Commute(\text{CNOT}_7(0,1,1,1),\text{CNOT}_8(0,0,1))$= TRUE, $\text{CNOT}_8(0,0,1)$ cannot be executed.

\item $Non$-$Execute(\text{CNOT}_2(1,0,0,0), \text{CNOT}_9(1,1,0,0))$= TRUE since the second condition of $Non$-$Execute$ function is satisfied.

\end{itemize}

The steps described above correspond to the first run of Algorithm~\ref{alg:mint} and lead to the $DQC$ of Figure~\ref{fig:QC2}. The next runs of the algorithm are depicted in Figures~\ref{fig:QC3} to \ref{fig:QC4}. There is a final run in the algorithm that results in an empty circuit and it is not shown.

For each configuration of global gates of Figure~\ref{fig:QC1}, Table~\ref{Table1} shows the minimum number of teleportations and their corresponding sequence. In this table, the general name $g$ for gates of $DQC$ is used. $g_k(C \slash T)$ implies that the control$\slash$target qubit of $g_k$ is teleportated to the other partition. The configuration used in the example is configuration \#9 in Table \ref{Table1}.

 In this example, the minimum number of teleportations, which is 4, occurs for the configuration \#24. This configuration corresponds to the $Config$-$Arr$$=$$\{11000\}$ where the first and second global gates are executed in $P_1$ and other global gates are executed in partition $P_0$.

As Table~\ref{Table1} shows, in the worst case, this example requires 10 teleportations while the optimizations of the proposed approach decrease this value to 4. Therefore, a remarkable improvement, i.e., $60\%$ is obtained for this small circuit by the proposed approach.
\begin{table}
\caption{The sequence and the number of minimum teleportations for different configurations of Figure~\ref{fig:QC1}.}
\label{Table1}
\begin{tabular}{|c|c|c|}
\hline
Configuration \#&The sequence of qubit&$n_t$\\
(in base 10)&teleportations&\\
\hline
0&$g_2(C), g_3(T), g_5(C)$&6\\
\hline
1&$g_2(C), g_3(T), g_5(C), g_9(T)$&8\\
\hline
2&$g_2(C), g_3(T), g_5(C), g_7(C), g_9(C)$&10\\
\hline
3&$g_2(C), g_3(T), g_5(C), g_7(C), g_9(T)$&10\\
\hline
4&$g_2(C), g_3(T), g_5(T), g_7(T)$&8\\
\hline
5&$g_2(C), g_3(T), g_5(T), g_7(T), g_9(T)$&10\\
\hline
6&$g_2(C), g_3(T), g_5(T), g_7(C), g_9(C)$&10\\
\hline
7&$g_2(C), g_3(T), g_5(T), g_7(C), g_9(C)$&10\\
\hline
8&$g_2(C), g_3(C), g_5(C)$&6\\
\hline
9&$g_2(C), g_3(C), g_5(C), g_9(T)$&8\\
\hline
10&$g_2(C), g_3(C), g_5(C), g_7(C), g_9(C)$&10\\
\hline
11&$g_2(C), g_3(C), g_5(C), g_7(C), g_9(T)$&10\\
\hline
12&$g_2(C), g_3(C), g_5(T), g_7(T)$&8\\
\hline
13&$g_2(C), g_3(C), g_5(T), g_7(T), g_9(T)$&10\\
\hline
14&$g_2(C), g_3(C), g_5(T), g_7(C), g_9(C)$&10\\
\hline
15&$g_2(C), g_3(C), g_5(T), g_7(C), g_9(T)$&10\\
\hline
16&$g_2(T), g_3(T), g_5(C)$&6\\
\hline
17&$g_2(T), g_3(T), g_5(C), g_9(T)$&8\\
\hline
18&$g_2(T), g_3(T), g_5(C), g_7(C), g_9(T)$&10\\
\hline
19&$g_2(T), g_3(T), g_5(C), g_7(C), g_9(T)$&10\\
\hline
20&$g_2(T), g_3(T), g_5(T), g_7(T)$&8\\
\hline
21&$g_2(T), g_3(T), g_5(T), g_7(T), g_9(T)$&10\\
\hline
22&$g_2(T), g_3(T), g_5(T), g_7(C), g_9(C)$&10\\
\hline
23&$g_2(T), g_3(T), g_5(T), g_7(C), g_9(T)$&10\\
\hline
24&$g_2(T), g_5(C)$&4\\
\hline
25&$g_2(T), g_5(C), g_9(T)$&6\\
\hline
26&$g_2(T), g_5(C), g_7(C) ,g_9(C)$&8\\
\hline
27&$g_2(T), g_5(C), g_7(C), g_9(T)$&8\\
\hline
28&$g_2(T), g_5(T), g_7(T)$&6\\
\hline
29&$g_2(T), g_5(T), g_7(T), g_9(T)$&8\\
\hline
30&$g_2(T), g_5(T), g_7(C), g_9(C)$&8\\
\hline
31&$g_2(T), g_5(T), g_7(C), g_9(T)$&8\\
\hline
\end{tabular}
\end{table}

\section{Conclusion and Future Works}

In this study, we presented an algorithm to optimize the number of teleportations for a distributed quantum system consisting of two spatially separated and long-distance quantum subsystems. As teleportation is a costly operation in quantum computation technologies, reducing the number of such operations is very important for designing distributed quantum computers.

As future works, extending the proposed approach to the case when the number of partitions is more than two can be considered. Moreover, an algorithmic approach for the quantum circuit partitioning can be used instead of considering a preassigned partitioning for the subsystems. This step can then be followed by the proposed approach for teleportation cost optimization. Finally, the used library of gates can be extended to include three-qubit quantum gates and hence the proposed algorithm should be modified accordingly.

\bibliographystyle{unsrt}
\bibliography{ref}

\begin{IEEEbiographynophoto}
{Mariam Zomorodi-Moghadam}
received her BSc. degree in computer engineering from
Iran University of Science and Technology, in 2002 and her MSc. degree in computer
engineering from Amirkabir University of Technology (Tehran Polytechnic), in 2007. She
received her PhD degree in computer engineering from Shahid Beheshti University, in 2014.
She is currently an Assistant Professor in the Department of Computer Engineering at
Ferdowsi University of Mashhad. Her current research interests include quantum computing,
reversible computing, embedded system design, and data mining.
\end{IEEEbiographynophoto}

\begin{IEEEbiographynophoto}
{Monireh Houshmand}
received the BS, MS, and PhD degrees in electrical engineering from Ferdowsi University of Mashhad, in 2005, 2007, and 2011 respectively. At present, she is an assistant professor of electrical engineering at Imam Reza International University, Mashhad, Iran. Her current research interests include distributed quantum computing, quantum error correction, and quantum cryptography.
\end{IEEEbiographynophoto}

\begin{IEEEbiographynophoto}
{Mahboobeh Houshmand}
received the BS, MS degrees in computer engineering from Ferdowsi University of Mashhad, in 2007 and 2010, respectively and the PhD degree from Amirkabir University of Technology in 2015. Presently, she is an assistant professor of computer engineering at Mashhad Branch, Islamic Azad University, Mashhad, Iran. Her current research interests are in the areas of distributed quantum computing, quantum logic synthesis, and quantum machine learning.
\end{IEEEbiographynophoto}
\end{document}